\documentclass[aip,rsi,11pt,a4paper,onecolumn]{revtex4-1}
\usepackage[utf8x]{inputenc}

\usepackage{dcolumn}
\usepackage{bm}
\usepackage{amsmath}
\usepackage{amsfonts}
\usepackage{amssymb}
\usepackage{graphicx}
\usepackage{color}

\begin{document}

\title{\(LC\)-Circuit Calorimetry}
\author{O. Bossen}
\author{A. Schilling}
\affiliation{Physik-Institut der Universit\"at Z\"urich, Winterthurerstrasse 190, 8057 Z\"urich, Switzerland}
\begin{abstract}
We present a new type of calorimeter in which we couple an unknown heat capacity with the aid of Peltier elements to an electrical circuit.
The use of an electrical inductance and an amplifier in the circuit allows us to achieve  autonomous oscillations, and the measurement of the corresponding resonance frequency makes it possible to accurately measure the heat capacity with an 
intrinsic statistical uncertainty that decreases as \( \sim t_\mathrm{m}^{ -3/2}\) with measuring time \(t_\mathrm{m}\), as opposed to a corresponding uncertainty 
\(\sim t_\mathrm{m}^{-1/2}\) in the conventional alternating current (a.c.) method to measure heat capacities. We have built a demonstration experiment 
to show the feasibility of the new technique, and we have tested it on a gadolinium sample at its transition to the ferromagnetic state.
\end{abstract}
\maketitle

\section{Introduction}

Calorimetry is an old field in experimental physics, and numerous experimental techniques have been developed for the accurate measurement of heat capacities
\cite{GmelinReview,*StewardCalReview,*ModCalReview}.
The frequently used a.c. technique invented by Corbino \cite{Corbino_AC_versuche,*Corbino_AC_erfindung} and modified by Sullivan and Seidel\cite{ACRediscovery}, for example,
measures a heat capacity by monitoring the temperature amplitude of a sample that is subject to an oscillating heating power , or, although less common, by measuring an oscillating heat current produced by a Peltier element \cite{ThermocoupleAC,*ModulatedBath}.
Such a.c. techniques have numerous advantages over other methods, e.g., heat-pulse, differential-thermal-analysis\cite{MarkCal} or relaxation techniques \cite{GmelinReview,*StewardCalReview,*ModCalReview}.
A certain robustness to noise and the possibility to use very small temperature amplitudes stems from the fact that the signal of interest can be separated from
the instrumental noise by using a frequency selective filter, such as a lock-in amplifier. To increase the total accuracy even further, the duration of a measurement \(t_{\mathrm{m}}\)
can be increased at will, and the resulting statistical uncertainty decreases as \( t_\mathrm{m}^{-1/2}\).\\
If a calorimeter could be built where the heat capacity is related to the frequency instead of the amplitude of an oscillation one would retain most of the advantages of the conventional
a.c. techniques, but the statistical uncertainty would decrease much more rapidly \cite{Rife_N32}, i.e., as \( t_\mathrm{m}^{-3/2}\).
Therefore, the accuracy of such a method would be mainly limited by the stability of the oscillator components, rather than by statistical constraints.\\
In the following we describe the realization of an oscillating thermo-electrical circuit using Peltier elements which makes an electrical inductor to effectively act as a ``thermal inductor``.
The heat capacity can be calculated from the measured resonance frequency of the circuit, and its statistical uncertainty drops according to \(t_{\mathrm{m}}^{-3/2}\) as expected.

\section{Operating principle of the thermo-electrical oscillator}
\begin{figure}
 \centering
 \includegraphics[width=6cm,keepaspectratio=true]{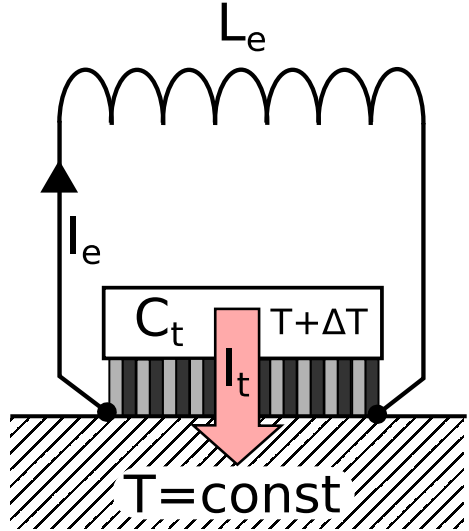}
 \caption{An electrical inductor \(L_{\mathrm{e}}\) is connected to a Peltier element that separates a heat capacity \(C_{\mathrm{t}}\) from a thermal bath \(T\). Changes in the electrical
current \(I_{\mathrm{e}}\) induce a temperature difference \(\Delta T = L_{\mathrm{e}} \dot{I}_{\mathrm{e}}/\alpha\) (where \(\alpha\) denotes the Seebeck coefficient) across the
Peltier element that counteracts the variation in the flow of heat \(\dot{I}_{\mathrm{t}}\).}
 \label{fig:PeltierCoil}
\end{figure}

It is known that the flow of heat \(Q_{\mathrm{t}}\) (with a corresponding heat
current \(I_{\mathrm{t}} = \dot{Q}_\mathrm{t}\)) between entities with a certain heat capacity \(C_{\mathrm{t}}\) through links with a finite
thermal conductivity can be mapped onto the flow of charge \(Q_\mathrm{e}\) (with the electrical current \(I_\mathrm{e} = \dot{Q}_{\mathrm{e}}\))
in electrical circuits that is composed of capacitors \(C_{\mathrm{e}}\) and resistors, respectively, where the temperature \(T\) in the thermal
case corresponds to the electrostatic potential \(V\) in the equivalent electrical circuit. The differential equations describing the time evolution of
\(Q_\mathrm{t}\left( t\right) \) or \(Q_\mathrm{e}\left( t\right) \), respectively, are of first order in time \(t\). However, a simple thermal analogue
of an electrical inductor involving a higher derivative
\(\ddot{Q}_{\mathrm{t}} = \dot{I}_{\mathrm{t}}\) does not exist, although the inclusion of terms including 
\(\dot{I}_{\mathrm{t}}\) is justified in certain rare cases \cite{Bosworth_t_induction,*McNelly_second_sound,*Weedy_thermal_ind,*Joseph_heatwave_review}.
Therefore it is impossible to realize autonomously oscillating thermal circuits by using only standard thermal elements.

In our experiment we connect an electrical inductor \(L_{\mathrm{e}}\) to a Peltier element as shown in Fig.\ \ref{fig:PeltierCoil}. If the heat
current \(I_{\mathrm{t}}\) through the Peltier element to the thermal bath varies with time, an associated electrical current \(I_{\mathrm{e}}\) in the circuit changes according to
\( \Pi \dot{I}_{\mathrm{e}} = \dot{I}_\mathrm{t}\) , where \(\Pi \) is the Peltier coefficient of the Peltier element. It induces a voltage drop \(L_{\mathrm{e}} \dot{I}_{\mathrm{e}}\)
across the inductor that opposes the changes in the heat current \(I_{\mathrm{t}}\) through the Peltier element via the Seebeck effect, i.e., a resulting
additional temperature difference \(\Delta T = L_{\mathrm{e}} \dot{I}_{\mathrm{e}}/ \alpha \) (where \( \alpha \) denotes the Seebeck coefficient) across the Peltier element is
imposed that counteracts the variation in the flow of heat, \(\dot{I}_{\mathrm{t}}\) . We can therefore define an
effective thermal inductance
\begin{equation}
L_{\mathrm{t}} := \frac{\Delta T}{ \dot{I}_{\mathrm{t}}} = \frac{L_{\mathrm{e}}}{\alpha \Pi} \;.
\label{eq:lelt}
\end{equation}
However, as a consequence of the significant losses produced by the non-zero thermal conductivity and electrical resistance
of the Peltier element, the idealized element shown
in Fig.\ \ref{fig:PeltierCoil} allows in general only for an overdamped variation of the heat current.
This obstacle can be avoided by including an amplifier with a suitable gain in the electrical circuit, and
autonomous oscillations can eventually be achieved in this way. In the ideal case of a
stationary oscillation with constant amplitude where all losses are exactly compensated by
the amplifying element, the electrical circuit behaves like a simple \(LC\) circuit with 
thermal inductance \( L_{\mathrm{t}} \) and a heat capacity \(C_{\mathrm{t}}\), or with an electrical inductance 
\( L_{\mathrm{e}} \) and an equivalent electrical capacitance

\begin{equation}
C_{\mathrm{e}} = \frac{Q_{\mathrm{e}}}{V} = \frac{\int I_{\mathrm{e}} \, \mathrm{dt}}{\alpha \Delta T}
= \frac{ C_{\mathrm{t}} \int I_{\mathrm{e}} \, \mathrm{dt} }{\alpha \int \Pi I_{\mathrm{e}} \, \mathrm{dt}}
= \frac{C_{\mathrm{t}}}{\alpha \Pi}
\label{eq:ctce}
\end{equation}
and with resonance frequency \(\omega^2 = 1/L_{\mathrm{t}}C_{\mathrm{t}} = 1/L_{\mathrm{e}}C_{\mathrm{e}}\). Therefore, the heat capacity can be derived from \( \omega \) 
according to
\begin{equation}
C_{\mathrm{t}} = \frac{\alpha \Pi}{L_{\mathrm{e}} \omega^2} \; ,
\label{Eq:CofOmega}
\end{equation}
notably without any need to quantify the thermal and electrical
losses in the Peltier element. To the best of our knowledge, the correspondences between an \(L_{\mathrm{e}}\) and a thermal analogue \(L_{\mathrm{t}}\),
or \(C_{\mathrm{t}}\) and a \(C_{\mathrm{e}}\) as expressed in Eqs. (\ref{eq:lelt}) and (\ref{eq:ctce}), respectively,
are novel and have also never been exploited in any calorimetric technique.

\section{Experimental setup}

\begin{figure*}
 \centering
 \includegraphics[width=12cm,keepaspectratio=true]{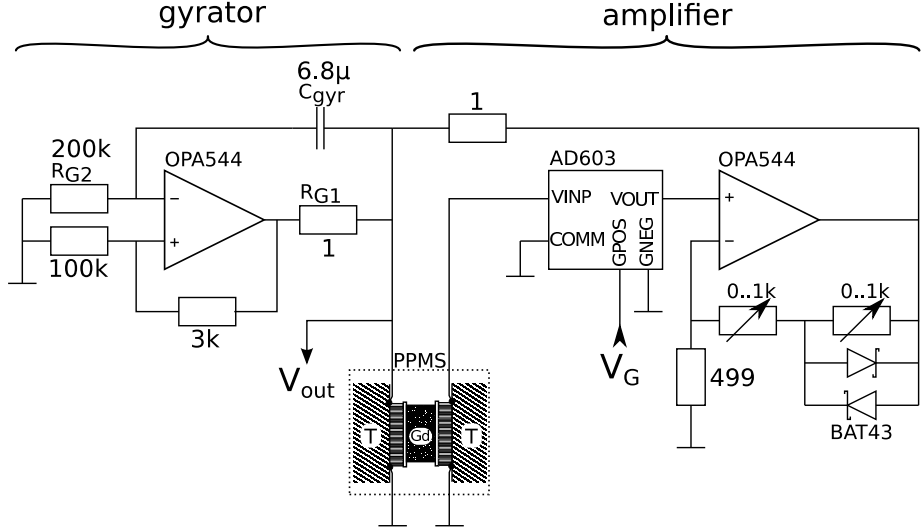}
 \caption{Circuit diagram showing the Peltier element system, the Gyrator to mimic a large electrical inductance, and the amplifier to compensate for thermal and electrical losses (see text). The voltage $V_\mathrm{G}$
controls the gain of the amplifier. The two Peltier elements have the same polarity, and the sample is attached to their ``top side``. The measured signal is tapped at \(V_{\text{out}}\).}
 \label{fig:Schaltplan}
\end{figure*}

\begin{figure*}
 \centering
 \includegraphics[width=12cm,keepaspectratio=true]{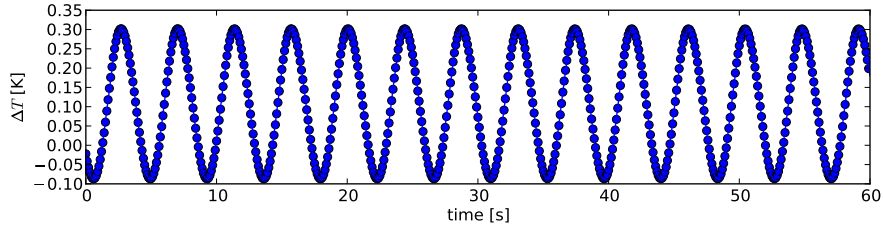}
 % single_oscillation.pdf: 1500x350 pixel, 100dpi, 38.10x8.89 cm, bb=0 0 1080 252
 \caption{Thermal oscillations of the \(LC\)-circuit. The solid line is a least-squares fit to a sinusoidal variation
of \(\Delta T\) in time around an offset value that is caused by lost heat dissipated in the Peltier elements.}
 \label{fig:single_oscillation}
\end{figure*}

We have chosen the circuit shown in Fig.\ \ref{fig:Schaltplan} containing two Peltier elements (MPC-D701, \emph{Micropelt Inc.}) based on \(\mathrm{Bi_2Te_3}\), with lateral dimensions of
\(\approx 3.5 \, \mathrm{mm}\) and a thickness of \(\approx 1 \, \mathrm{mm}\). While the left element acts, together with the sample to be measured, as the equivalent electrical capacitance
as described above, the right element is used to compensate for thermal and electrical losses and to maintain a constant amplitude of the temperature oscillation. For this purpose, the
voltage across the right element is amplified and then fed onto the electrical contacts of the left element. To achieve a stable oscillation of the desired amplitude, the gain of the
amplifier can be controlled using a voltage-controlled gain amplifier (AD630, \emph{Analog Devices}). An additional clamp with Schottky diodes (BAT43) and two potentiometers  forms a nonlinear
resistor that reduces the gain at large oscillation amplitudes and therefore forces the circuit to self-stabilize.\\
For a design heat capacity \(C_{\mathrm{t}} \approx 20\, \mathrm{mJ/K}\) to be measured near room temperature (which corresponds to \(2.87\cdot 10^{-4}\, \mathrm{mol}\) of a Dulong-Petit solid), and using   \(\alpha \approx 29\, \mathrm{mV/K}\) and
\(\Pi \approx \alpha T \approx 8.4\, \mathrm{W/A}\) for the used Peltier elements, the equivalent electrical capacitance becomes \(C_{\mathrm{e}} \approx 82 \,\mathrm{mF}\). Depending on
the heat capacities and the thermal conductivities involved in thermal part of the system, the frequency of the temperature oscillation must be small enough to allow for a
quasi-stationary thermal equilibrium. If we choose   \(f=\omega / 2 \pi \approx 0.18 \,\mathrm{Hz}\) we  need an inductance \(L_{\mathrm{e}} \approx 10 \,\mathrm{H}\) which would require impractically
large passive elements. Therefore we replaced the inductance by a gyrator, the circuit diagram of which is also shown in Fig.\ \ref{fig:Schaltplan}. The effective induction of the gyrator is given by \cite{Gyrator}
\(L_{\mathrm{e}} = (R_{\mathrm{P}} + R_{\mathrm{G1}})R_{\mathrm{G2}} C_{\mathrm{gyr}}\), where the internal resistance \( R_{\mathrm{P}} \approx 6.5\, \mathrm{\Omega}\)  of the Peltier element has to be
 included here. With  \( R_{\mathrm{G1}}= 1\, \mathrm{\Omega}\)  , \(  R_{\mathrm{G2}}= 200\, \mathrm{k\Omega}\)  and  \( C_{\mathrm{gyr}}= 6.8\, \mathrm{\mu F}\) we indeed obtain \(L_{\mathrm{e}} \approx 10 \,\mathrm{H}\).

\section{Test measurements}

\begin{figure}
 \centering
 \includegraphics[width=8cm,keepaspectratio=true]{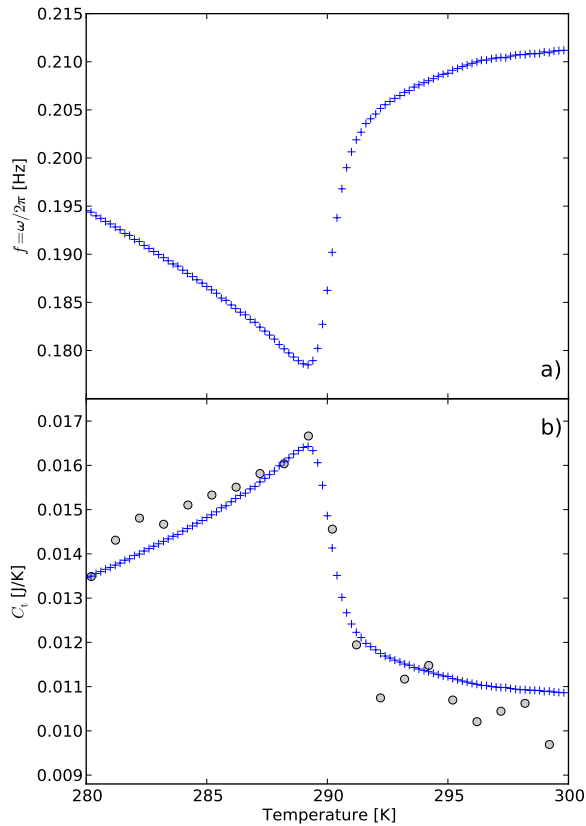}
 
 \caption{a) Measured frequency of the oscillator %b) magnetic moment
and b) heat capacity of a Gd sample, as functions of temperature.
Crosses represent the data that we obtained by the present thermal \(LC\)-method, while circles denote the values from a measurements on the same sample using a commercial heat capacity
system (see text).}
 \label{fig:TwoParameters}
\end{figure}

A test sample (45.2 mg or \(2.87\cdot 10^{-4}\, \mathrm{mol}\) of gadolinium with 99.99\% purity that had been previously annealed at \(T = 1123 \, \mathrm{K}\) in vacuum to reduce lattice strains \cite{GdFieldDependent}) was sandwiched
between the two Peltier elements and placed into a commercial PPMS system (Physical Property Measurement System, \emph{Quantum Design}) for control of the base temperature. The gain of the
amplifier was chosen in a way to achieve temperature oscillations with an amplitude of the order of \(0.2 \,\mathrm{K}\). For each data point, the oscillating voltage was read out during 1200 s with
a 10 Hz-triggered Keithley 2001 Multimeter in equal time intervals of 0.1\ s (see Fig.\ \ref{fig:single_oscillation}), and the resulting data were then least-squares fitted to obtain the oscillation frequency (see Fig.\ \ref{fig:TwoParameters}a).
To compensate for the heat capacity of the empty system (\(\approx 15\%\) of the total heat capacity), corresponding measurements have been done without the sample installed.
The resulting values of the heat capacity are shown in Fig.\ \ref{fig:TwoParameters}b, together with heat-capacity data obtained on the same sample using the commercial heat-capacity option of the PPMS system
and with temperature steps of \( \approx 0.5\, \mathrm{K}\). While the heat-capacity data using our thermal \(LC\)- method are neither smoothed nor further averaged, the PPMS data were averaged over 6 individual 
measurements. The data-acquisition time for each of the \(LC\)- method data points was 1200 s, while collecting an averaged PPMS data point took about one hour of time, and the temperature amplitudes for both techniques were comparable (0.3 K - 0.5 K).
The transition of the Gd sample to the ferromagnetic state takes place at the Curie temperature \(T_{\mathrm{C}} \approx 290 \mathrm{K}\), and the absolute values for the heat capacity agree very well with the PPMS data (see Fig.\ \ref{fig:TwoParameters}b).

\section{Accuracy considerations}

\begin{figure}
 \centering
 \includegraphics[width=8cm,keepaspectratio=true]{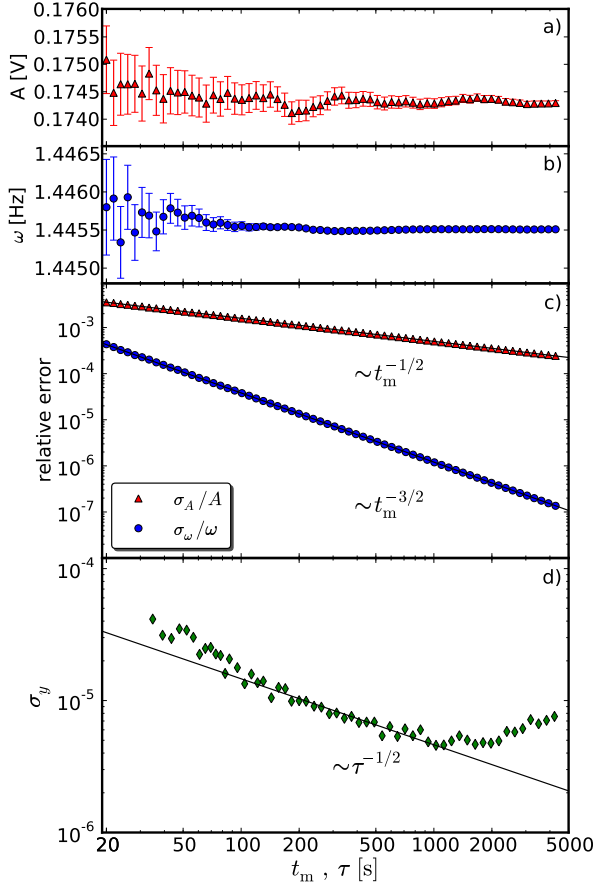}
 \caption{a) Amplitudes and b) frequencies for a selected data point together with the respective
standard deviations from the Levenberg-Marquardt fit, as functions of measurement time \(t_{\mathrm{m}}\).
c) The respective relative standard deviations on a log-log scale to illustrate the
power-law behavior discussed in the text  Solid lines are fits according to \( t^{-1/ 2}\) for \(\sigma_A/A  \) and to \(t^{-3 / 2}\) for the \(\sigma_\omega /\omega  \) data, respectively.
d) The overlapping Allan deviation \(\sigma_y\), plotted against the observation time $\tau$. The solid line illustrates the approximate \(\tau^{-1/ 2}\) dependence of \(\sigma_y\) (see text).}
 \label{fig:comparison}
\end{figure}

The accuracy of any heat-capacity measurement is limited by systematic and statistical uncertainties. Sources of systematic errors (such as finite instrumental resolution, limited accuracy by a large background signal,
errors in the temperature calibration etc.) are common to all techniques for measuring a heat capacity, and we shall not further discuss these shortcomings here. The main advantage of the present
thermal \(LC\)-circuit technique is that it significantly reduces the statistical uncertainty with increasing measuring time.

The statistics behind the problem of estimating the amplitude \(A\), the frequency \(\omega\) and the phase \(\varphi\) of an oscillation \(A \sin(\omega t +\varphi )\) out of a set of \(N\) samples of finite length \(T\)
has been analyzed in detail in Ref.\ \onlinecite{Rife_N32}. Assuming a white noise with variance \(\sigma^2\), the resulting relative standard deviation \(\sigma_{A}/A\) 
in the amplitude \(A\) from a measurement of \(N\) oscillation cycles with length \(T = 2\pi /\omega\) is
\begin{equation}
\frac{\sigma_A}{A}  \geq \frac{\sigma} {A \sqrt{N}} \;. 
\end{equation}

In a conventional a.c. technique to measure heat capacities, the relative standard deviation in the heat
capacity \(C_{\mathrm{t}}\) from the mean value is proportional to the respective standard deviation in the measured amplitude \(A\) of a temperature oscillation \cite{Corbino_AC_erfindung},
and therefore the statistical uncertainty in \(C_{\mathrm{t}}\) decreases with \(t_{\mathrm{m}}^{-1/2}\) where \(t_\mathrm{m} = NT\) is the total time to take such a
measurement (in all other conventional methods, the statistical uncertainty also decreases as \(N^{-1/ 2}\), where \(N\) is the number of measurements to calculate the average of a data point).

The corresponding relative standard deviation \(\sigma_{\omega}/\omega\) in the frequency \(\omega\) , however, varies with \(N\) according to
\begin{equation}
 \frac{\sigma_\omega}{\omega}  \geq \frac{\sqrt{3} \sigma}{\pi A \sqrt{N(N^2 -1)}}
\label{eqn:exact_error_formula}\; ,
\end{equation}
i.e., it decreases with \( N^{-3/2}\) (or \( t_{\mathrm{m}}^{-3/2}\) ) for large \(N\). As the heat capacity in our technique is proportional to \( \omega^{- 2}\) (see Eq.
\ref{Eq:CofOmega}), the relative standard deviation \(\sigma_{C_{\mathrm{t}}} /C_{\mathrm{t}}  \approx 2 \cdot \sigma_\omega /\omega  \) also decreases with \(N\) in the same fashion.
In Fig.\ \ref{fig:comparison} we show the amplitudes and the frequencies for a selected data point, together with the respective relative standard deviations, as functions of the measurement time \(t_{\mathrm{m}}\) which is proportional to
the number of oscillation cycles \(N\). The standard deviations \(\sigma_{A}\) and \(\sigma_\omega \) have been taken from a standard Levenberg-Marquardt routine used for least-squares fitting the
experimental data. The \(\sigma_A/A \) data decrease relatively slowly with \(N^{-1/ 2}\) , while the \(\sigma_\omega / \omega \) data (and along with them
\(\sigma_{C_{\mathrm{t}}} /C_{\mathrm{t}} \) ) drop much faster with \(N^{-3 / 2}\) as expected. Therefore, very high accuracies for \(C_{\mathrm{t}}\) can, in principle, be achieved in a comparably very short measuring time.
For example, to reduce the statistical uncertainty in \(C_{\mathrm{t}}\) by a factor of 10, conventional methods require a 100 times longer duration of a measurement, while the thermal \(LC\) method
needs only a \(\approx 5\) times longer data acquisition time. As both \(\sigma_A/A\) and \(\sigma_\omega /\omega \) are proportional to the same quantity \(\sigma /A\), there is no hidden drawback
in the equation \ref{eqn:exact_error_formula} that would put this tremendous statistical advantage into perspective.

However, in order to tap the full potential of this technique it is essential that the oscillation frequency \(\omega \) is sufficiently stable during the time of the measurement so that the
potentially minuscule statistical uncertainty is not impaired by extrinsic fluctuations, e.g., by an insufficient stabilization of the temperature of the sample or by an unwanted variability in
the electronic circuit. A standard measure for the stability of an oscillator is provided by the Allan deviation \(\sigma_y\) \cite{AllanPaper}. Assuming white noise in the oscillator frequency \(\omega\),
for example, \(\sigma_y\) is expected to drop as \(\tau^{-1/2}\), where \(\tau\) denotes the time over which the frequency is observed. At large observation times \(\sigma_y\) can saturate or even increase again as soon as other 
sources of noise (Flicker noise, random-walk noise) become dominant \cite{NISTADEV}. In Fig. \ref{fig:comparison}d we have plotted the overlapping Allan deviation as a function of observation time \(\tau\) on a log-log scale.
While \(\sigma_y\) initially drops approximately with \(\tau^{-1/2}\), is saturates in our experiment around \(\sigma_y\approx 4\cdot 10^{-6}\) for \(\tau \approx 1000\,\mathrm{s}\) an increases for larger vales of \(\tau\).
In this sense we interpret the minimum in \(\sigma_y\) as an estimate for the achievable uncertainty in \(\omega\) for a given set of experimental data, while the \(\sigma_\omega/\omega\) values from Eq. \ref{eqn:exact_error_formula}
and shown in Fig. \ref{fig:comparison}c represent the statistical uncertainty assuming an infinitely stable frequency with white noise only in the temperature amplitude. In our demonstration experiment, no special care has been taken to minimize
extrinsic sources for variations in \(\omega\). Such sources of error are the decisive factors that are limiting the maximum achievable accuracy
in the present thermal \(LC\) technique but they are to a large extent controllable, while the intrinsic statistical uncertainty should play a significantly less important role than in conventional techniques to measure heat capacities.

\section{Conclusions}

We have described a method where a heat capacity is coupled to an electrical circuit with the aid of Peltier elements, 
which makes an electrical inductor to effectively act as a ``thermal inductor``. The use of an amplifier in the circuit allows
us to achieve autonomous oscillations, and the measurement of the corresponding resonance frequency makes it possible to calculate the heat capacity. While the statistical uncertainty in
conventional techniques to measure heat capacities scales according to \(t_{\mathrm{m}}^{-1/ 2}\) with measuring time \(t_{\mathrm{m}}\), the respective uncertainty in the present thermal \(LC\) technique decreases
significantly faster, i.e., with \(t_{\mathrm{m}}^{ -3 / 2}\), and is ultimately limited by the stability of the thermal and electronic components used in the experiment .

\begin{acknowledgments}
This work was supported by the \emph{Schweizerische Nationalfonds zur F\"orderung der Wissenschaftlichen Forschung}, Grant. No. 20-131899 
\end{acknowledgments}

%\bibliography{Citations} 
%merlin.mbs aipnum4-1.bst 2010-07-25 4.21a (PWD, AO, DPC) hacked
%Control: key (0)
%Control: author (8) initials jnrlst
%Control: editor formatted (1) identically to author
%Control: production of article title (-1) disabled
%Control: page (0) single
%Control: year (1) truncated
%Control: production of eprint (0) enabled
%

\end{document}